\documentclass[sigconf,nonacm]{acmart}
\usepackage{multirow}
\usepackage{hyperref} %
\usepackage[english]{babel} 
\usepackage{caption}
\usepackage{subcaption}
\usepackage{booktabs}
\usepackage{enumitem}

\usepackage{listings}

\usepackage[most]{tcolorbox}   %
\usepackage{xcolor}            %
\usepackage{fvextra}           %

\usepackage{makecell}
\usepackage{tabularx}
\usepackage{url}
\usepackage{pifont}
\usepackage{acronym}

\newcommand{\cmark}{\ding{51}}  %
\newcommand{\xmark}{\ding{55}}  %

\DefineVerbatimEnvironment{myverbatim}{Verbatim}{
  breaklines=true,
  breakanywhere=true
}

\makeatletter
\def\tcb@cnt@datalistautorefname{Listing}
\makeatother

\newtcolorbox[auto counter]{datalist}[2][]{%
  enhanced,
  breakable,                             %
  fonttitle=\bfseries\small,
  fontupper=\small,
  colback=gray!10,                       %
  colframe=black,                        %
  fonttitle=\bfseries,                   %
  title={Listing \thetcbcounter. #2},       %
  title after break={Listing \thetcbcounter. #2 (continued)}, %
  rounded corners,                         %
  boxrule=0.5pt,                         %
  top=5pt, bottom=5pt, left=5pt, right=5pt, %
before=\par\vspace{10pt}\noindent,     %
  after=\par,               %
  before upper={\setlength{\parindent}{0pt}}, %
  width=\textwidth,                      %
  #1                                     %
}

\definecolor{kw}{HTML}{0B3D91}      %
\definecolor{str}{HTML}{8B2E2E}     %
\definecolor{cmt}{HTML}{2F7D32}     %
\definecolor{num}{HTML}{666666}     %
\definecolor{rule}{HTML}{999999}    %

\newcounter{prompt}[section]
\renewcommand{\theprompt}{\thesection.\arabic{prompt}}

\newtcolorbox[use counter=prompt]{promptbox}[1][]{%
  colback=gray!10,              %
  colframe=black!40,            %
  title=Prompt~\theprompt,      %
  label=prompt:\theprompt,
  fonttitle=\bfseries\small,    %
  fontupper=\scriptsize\ttfamily\color{black!70},
  boxrule=0.4pt,                %
  arc=2mm,                      %
  left=4pt,                     %
  right=4pt,
  top=4pt,
  bottom=4pt,
  before upper={\vspace{2pt}},
  after upper={\vspace{2pt}},
  breakable,
  enhanced,
  #1
}

\usepackage[ruled,vlined,linesnumbered]{algorithm2e}
\usepackage{amsmath}
\DeclareRobustCommand{\A}[1]{\textbf{A#1}}                %

\SetCommentSty{mycommfont}

\makeatletter
\renewcommand{\sectionautorefname}{\S\@gobble}
\renewcommand{\subsectionautorefname}{\S\@gobble}
\renewcommand{\subsubsectionautorefname}{\S\@gobble}
\renewcommand{\figureautorefname}{Fig.\@gobble}
\renewcommand{\tableautorefname}{Tab.\@gobble}
\renewcommand{\appendixautorefname}{App.\@gobble}
\newcommand{\eqabbr}{Eq.}

\renewcommand{\equationautorefname}{\eqabbr\@gobble}
\makeatother

\AtBeginDocument{
\definecolor{patriarch}{rgb}{0.5, 0.0, 0.5}
\definecolor{acmblue}{cmyk}{1, 0.1, 0, 0.1}
\definecolor{acmdarkblue}{cmyk}{1, 0.2, 0, 0.15}
\hypersetup{
  colorlinks,
  citecolor=patriarch,
  linkcolor=patriarch,
  urlcolor=acmdarkblue
}
}

\acrodef{poc}[PoC]{Proof-of-Concept}
\acrodef{asr}[ASR]{Attack Success Rate}
\acrodef{mcp}[MCP]{Model Context Protocol}

\usepackage{cleveref}

\AtBeginDocument{%
  }

\begin{document}

\title{When MCP Servers Attack: Taxonomy, Feasibility, and Mitigation}

\author{Weibo Zhao}
\email{weibo.zhao@u.nus.edu}
\affiliation{%
  \institution{National University of Singapore}
  \country{Singapore}
}

\author{Jiahao Liu}
\email{jiahao99@comp.nus.edu.sg}
\affiliation{%
  \institution{National University of Singapore}
  \country{Singapore}
}

\author{Bonan Ruan}
\email{r-bonan@comp.nus.edu.sg}
\affiliation{%
  \institution{National University of Singapore}
  \country{Singapore}
}

\author{Shaofei Li}
\email{lishaofei@pku.edu.cn}
\affiliation{%
  \institution{Peking University}
  \city{Beijing}
  \country{China}
}

\author{Zhenkai Liang}
\email{liangzk@comp.nus.edu.sg}
\affiliation{%
  \institution{National University of Singapore}
  \country{Singapore}
}
\renewcommand{\shortauthors}{Zhao et al.}

\begin{abstract}
\ac{mcp} servers enable AI applications to connect to external systems in a plug-and-play manner, but their rapid proliferation also introduces severe security risks.
Unlike mature software ecosystems with rigorous vetting, MCP servers still lack standardized review mechanisms, giving adversaries opportunities to distribute malicious implementations.
Despite this pressing risk, the security implications of MCP servers remain underexplored.
To address this gap, we present the first systematic study that treats MCP servers as active threat actors and decomposes them into core components to examine how adversarial developers can implant malicious intent.
Specifically, we investigate three research questions: (i) what types of attacks malicious MCP servers can launch, (ii) how vulnerable MCP hosts and Large Language Models (LLMs) are to these attacks, and (iii) how feasible it is to carry out MCP server attacks in practice.
Our study proposes a component-based taxonomy comprising twelve attack categories.
For each category, we develop \ac{poc} servers and demonstrate their effectiveness across diverse real-world host–LLM settings. 
We further show that attackers can generate large numbers of malicious servers at virtually no cost. 
We then test state-of-the-art scanners on the generated servers and found that existing detection approaches are insufficient.
These findings highlight that malicious MCP servers are easy to implement, difficult to detect with current tools, and capable of causing concrete damage to AI agent systems.
Addressing this threat requires coordinated efforts among protocol designers, host developers, LLM providers, and end users to build a more secure and resilient MCP ecosystem.
\end{abstract}


\maketitle

\section{Introduction}\label{sec:introduction}

As LLMs become increasingly powerful, there is growing interest in enabling them to access up-to-date information and perform real-world actions.
Early solutions to this integration problem include custom API connectors and agent frameworks such as LangChain~\citep{langchain2025} and AutoGen~\citep{microsoft_autogen_2025}. 
While effective, these approaches typically require substantial developer effort.
The \ac{mcp}~\cite{mcp_introduction} provides an alternative by defining a standardized interface for resource discovery and tool invocation, allowing LLMs to dynamically access the external environment at runtime with minimal integration effort.
\ac{mcp} follows a client–server architecture where the \ac{mcp} host, typically an AI application, employs \ac{mcp} clients to establish connections to \ac{mcp} servers, which are wrappers for functions and data.
For example, an agent can connect to a local server to perform file system operations, and to a remote server to access email.
The \ac{mcp} architecture allows hosts and servers to be implemented independently, and the protocol design simplifies host-server integration by requiring only server registration on the host side.

The open, model-agnostic nature of MCP has fostered an active community.
By August 2025, over 16K \ac{mcp} servers were publicly available online~\cite{mcp_so_website}.
While major service providers such as Alibaba~\cite{alibaba2025opsmcp}, Google~\cite{google2025cloudrunmcp}, and Oracle~\cite{oracle2025sqlclmcp} have released official \ac{mcp} servers, the vast majority of servers are community-developed.
This rapid expansion comes with significant security risks. 
Unlike mobile app ecosystems, which benefit from mature vetting processes and centralized app stores, the \ac{mcp} ecosystem still lacks standardized oversight. 
On one hand, mainstream open-source platforms like GitHub~\cite{github2025} and npm~\cite{npm2025} allow anyone to publish \ac{mcp} servers freely.
On the other hand, even dedicated \ac{mcp} server hosting sites, such as Smithery.ai~\cite{smitherySmitheryExtend}, MCP.so~\cite{mcp_so_website}, and Glama~\cite{glamaModelContext}, have yet to establish robust auditing pipelines~\cite{song2025protocolunveilingattackvectors}. 
This opens the door for attackers to publish malicious \ac{mcp} servers disguised as useful tools, claiming advanced capabilities to lure users into installation.

These gaps highlight the urgent need for a deep understanding and systematic analysis of the threats brought by \ac{mcp} servers.
Several studies have begun to investigate \ac{mcp} server security~\cite{hou2025mcp_landscape_threats_directions,narajala2025enterprisegradesecuritymodelcontext,jing2025mcipprotectingmcpsafety,song2025protocolunveilingattackvectors,fang2025identifymitigatethirdpartysafety,wang2025mpmapreferencemanipulationattack}.
Some analyzed \ac{mcp}-related threats and proposed different taxonomies~\cite{hou2025mcp_landscape_threats_directions,narajala2025enterprisegradesecuritymodelcontext,jing2025mcipprotectingmcpsafety}.
Others introduced concrete attack vectors and conducted experiments on state-of-the-art LLMs~\cite{song2025protocolunveilingattackvectors,fang2025identifymitigatethirdpartysafety,wang2025mpmapreferencemanipulationattack}.
However, to the best of our knowledge, no prior work has systematically examined the security threats posed by malicious \ac{mcp} servers from the perspective of their components.
In this work, we address this gap by investigating the following  research questions:

\begin{itemize}[leftmargin=12pt]
    \item \textbf{RQ1:} What kind of attacks can malicious \ac{mcp} servers launch?
    \item \textbf{RQ2:} How vulnerable are \ac{mcp} hosts and LLMs to attacks from malicious \ac{mcp} servers?
    \item \textbf{RQ3:} How feasible are these attacks in practice, in terms of implementation and detection?
\end{itemize}

To answer these research questions, we first conduct a component-based \ac{mcp} server attack analysis ({\autoref{sec:taxonomy}}), and evaluate these attacks against various \ac{mcp} hosts and LLM models ({\autoref{sec:eval_attacks}}).
We further investigate the difficulty of attack implementation and detection bypassing ({\autoref{sec:eval_detection}}) to measure the real-world attack feasibility.

Based on the analysis and experiment results, we obtain the following key findings:
(1) Malicious \ac{mcp} servers are able to launch 12 categories of attacks, which can be organized into a component-based taxonomy.
Each category in the taxonomy encompasses multiple fine-grained attack types, which can lead to severe consequences such as system compromise, manipulation of LLMs and users, or client denial of service.
(2) The cross host–LLM evaluations demonstrate that the attacks in our taxonomy are effective even against advanced host–LLM combinations.
The experiments further reveal that the success of \ac{mcp} server attacks depends jointly on host design, LLM safety training, and user awareness.
(3) Malicious \ac{mcp} servers are easy to generate in practice, and existing \ac{mcp} server scanners are insufficient to detect them.
Collectively, these findings highlight the urgency of addressing \ac{mcp} server security and the necessity of considering malicious \ac{mcp} servers as a primary threat.
We discuss potential solutions and call for broader attention from the community, as securing the \ac{mcp} ecosystem requires coordinated efforts across multiple stakeholders, including \ac{mcp} protocol designers, \ac{mcp} host developers, LLM providers, and end users ({\autoref{sec:discuss}}).

In summary, this paper makes the following contributions:

\begin{itemize}[leftmargin=12pt]
    \item \textbf{Component-based taxonomy of \ac{mcp} server attacks.} We systematically analyze malicious \ac{mcp} servers from the perspective of their internal components, and develop a taxonomy covering 12 distinct attack categories. This taxonomy provides a structured foundation for understanding the attack surface of \ac{mcp} servers.

    \item \textbf{Empirical attack evaluation across hosts and LLMs.} We conduct extensive experiments to evaluate the effectiveness of identified attacks against diverse \ac{mcp} hosts and LLM models. Our results show that attacks are broadly effective and that their success depends on both \ac{mcp} host design and LLM safety alignment.
    
    \item \textbf{Feasibility analysis of real-world attacks.} We investigate the ease of implementing malicious \ac{mcp} servers, as well as their ability to bypass current detection mechanisms. We find that attacks are not only practical but also remain largely undetectable by existing auditing tools.
\end{itemize}

\section{Background}
\label{sec:background}

\subsection{Model Context Protocol}

\begin{figure}[t]
    \centering
    \includegraphics[width=\linewidth]{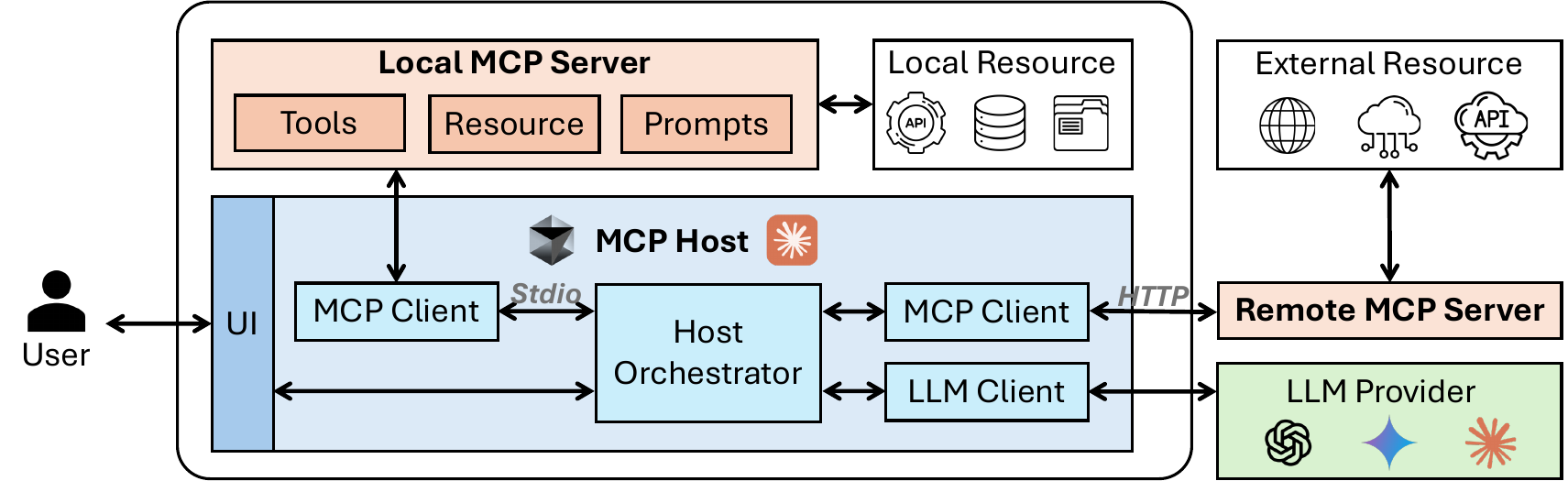}
    \caption{MCP architecture overview.}
    \label{fig:mcp_overview}
\end{figure}

\ac{mcp} is an open protocol that standardizes how LLMs interact with external data and tools~\citep{anthropic2024_introducing_mcp,mcp_introduction}.
~\autoref{fig:mcp_overview} illustrates the overview of MCP architecture, including three distinct components: \textbf{host}, \textbf{client}, and \textbf{server}.
The \ac{mcp} host is an LLM-based application, such as Claude Desktop~\cite{anthropic2025claudedesktop}, Cursor~\cite{cursor2025agents}, or Windsurf~\cite{windsurf2025}. 
Within a host, each MCP client establishes and maintains a one-to-one connection with an MCP server. 
The \ac{mcp} host may manage multiple such clients to support multiple servers.
The \ac{mcp} client handles message transport, tool discovery, tool invocation, and resource access on behalf of the host application.
The \ac{mcp} server provides the client with context, tools, and prompts, thereby extending the LLM with outward-facing capabilities~\cite{mcp_architecture}.

\subsection{MCP Servers}
MCP servers can be developed independently of host applications and easily plugged into different hosts, and thus have been widely adopted~\cite{mcp_so_2025,mcp_github_servers_2025,lin2025large}.
MCP servers can be categorized into local and remote servers based on their deployment model, as illustrated in \autoref{fig:mcp_overview}.
A local server runs on the same machine as the MCP host, allowing the LLM to access local data and services.
Local servers can be set up non-persistently using tools like \texttt{npx} or \texttt{uvx}, or installed persistently via \texttt{git clone}, \texttt{npm install}, or Docker.
In contrast, remote servers run on a different machine than the MCP host and typically provide access to remote or cloud-based resources.
An MCP server can be decomposed into six components based on distinct responsibilities: \textit{metadata, configuration, initialization logic, tools, resources,} and \textit{prompts}.
Every server must include metadata, configuration, and initialization logic in order to integrate with the host and operate normally. 
Tools, resources, and prompts are optional and depend on the services a server intends to provide.

Specifically, \textit{metadata} refers to the information about an MCP server that is intended for human or system consumption, including descriptive and structural types~\cite{nist2016cyberthreatsharing}.
Descriptive metadata is natural language content found in the documentation, README files, or promotional websites of the MCP server.
Structured metadata, typically represented in JSON format, appears in the server’s source code and the host’s config file, and includes fields such as name, version, and authorization settings.
\textit{Configuration} refers to the information the user and the host needs to use the server, typically provided by the server developer.
Server configuration encompasses launch configuration and connection configuration.
Launch configuration includes the executable paths, command-line arguments, and environment variables needed to start the server process.
Connection configuration is the information the MCP host requires to locate and connect to the server, such as a URL or port.
\textit{Initialization logic} refers to the code that enables the MCP server to start and integrate with the host. 
It typically includes dependency initialization, environment configuration, transport setup, and connection procedure.
Initialization logic can be regarded as the groundwork of an MCP server. 

Regarding the optional components, \textit{tools} are executable functions registered by an MCP server and exposed to the LLM as invokable endpoints. 
Each tool exposed by an MCP server comprises three components: the tool’s metadata, its internal logic, and its return value. 
The tool metadata includes: a name that identifies the tool; a human-readable description that LLM can understand; an input schema that specifies parameter names, types, descriptions and constraints using JSON Schema; and, optionally, a set of annotations that describe tool behavior.
\textit{Resources} are the standardized interface through which servers expose data to users and LLMs.
Each resource is uniquely identified by a URI and includes metadata such as a human-readable name, an optional description, a MIME type indicating content format, and optionally a size in bytes. 
\textit{Prompts} are reusable templates that a server provides to facilitate user interaction with the LLM.
Each prompt specifies a name, description, and a list of dynamic arguments.

\subsection{Limitations of Existing Work}

As MCP gains wider attention and adoption, recent research has begun to discuss its security challenges.
Hou et al.~\cite{hou2025mcp_landscape_threats_directions} highlight potential security risks associated with the creation, operation, and update phases of MCP servers, including installer spoofing, sandbox escape, and redeployment of vulnerable versions. 
From the perspective of the primary roles in MCP, Narajala and Habler~\cite{narajala2025enterprisegradesecuritymodelcontext} identify several categories of MCP security threats: MCP server threats, MCP client threats, MCP host environment threats, data sources and external resources threats, tool-related threats, and prompt-related threats.
Jing et al.~\cite{jing2025mcipprotectingmcpsafety} categorize risks in MCP interactions into five dimensions: Stage, Source, Scope, Type, and the corresponding MAESTRO layers. 
Based on their threat analysis, they introduce an enhanced protocol, MCIP, and develop an MCP interaction dialogue dataset, MCIP-Bench. 
Evaluations on MCIP-Bench show that existing LLMs have weak safety awareness in MCP scenarios.
Guo et al.~\cite{guo2025systematicanalysismcpsecurity} present the MCP Attack Library (MCPLIB), a unified, plugin-based framework for simulating and evaluating MCP attacks. They catalog 31 attacks across four families: direct tool injection, indirect tool injection, malicious-user attacks, and LLM-inherent attacks.

\begin{table}[t]
\centering
\small
\caption{Comparison of MCP security studies with this work.}
\resizebox{\linewidth}{!}{%
\begin{tabular}{lccccc}
\toprule
\textbf{} & \textbf{Hou~\cite{hou2025mcp_landscape_threats_directions}} & \textbf{Jing~\cite{jing2025mcipprotectingmcpsafety}} & \textbf{Narajala~\cite{narajala2025enterprisegradesecuritymodelcontext}} & \textbf{Guo~\cite{guo2025systematicanalysismcpsecurity}} & \textbf{Our work} \\
\midrule
Cross-LLM Attack Evaluation   & \xmark & \cmark & \xmark & \cmark & \cmark \\
Cross-Host Attack Evaluation  & \xmark & \xmark & \xmark & \xmark & \cmark \\
Scanner Effectiveness Evaluation          & \xmark & \xmark & \xmark & \xmark & \cmark \\
Artifact Availability                     & \xmark & \cmark & \xmark & \xmark & \cmark \\
\bottomrule
\end{tabular}}
\label{tab:literature_comparison}
\end{table}

While prior works provide valuable insights, none have systematically and empirically examined the threats posed by malicious MCP servers. 
In particular, as shown in \autoref{tab:literature_comparison}, empirical evaluation in existing work remains limited. 
In this paper, we address this gap through systematic analysis and extensive experiments. 
We analyze the internal structure of MCP servers and propose a component-based taxonomy of attacks. 
For each of the 12 attack categories in our taxonomy, we implement concrete \ac{poc} servers and conduct evaluations that goes beyond existing LLM-focused testing by exploring diverse host–LLM combinations. 
The result demonstrate that all identified MCP server attacks are practical and impactful; moreover, existing scanners fail to provide sufficient protection, and adversaries can easily implement malicious servers with minimal effort.

\section{MCP Server Attack Taxonomy}
\label{sec:taxonomy}

\begin{figure}
    \centering
    \includegraphics[width=\linewidth]{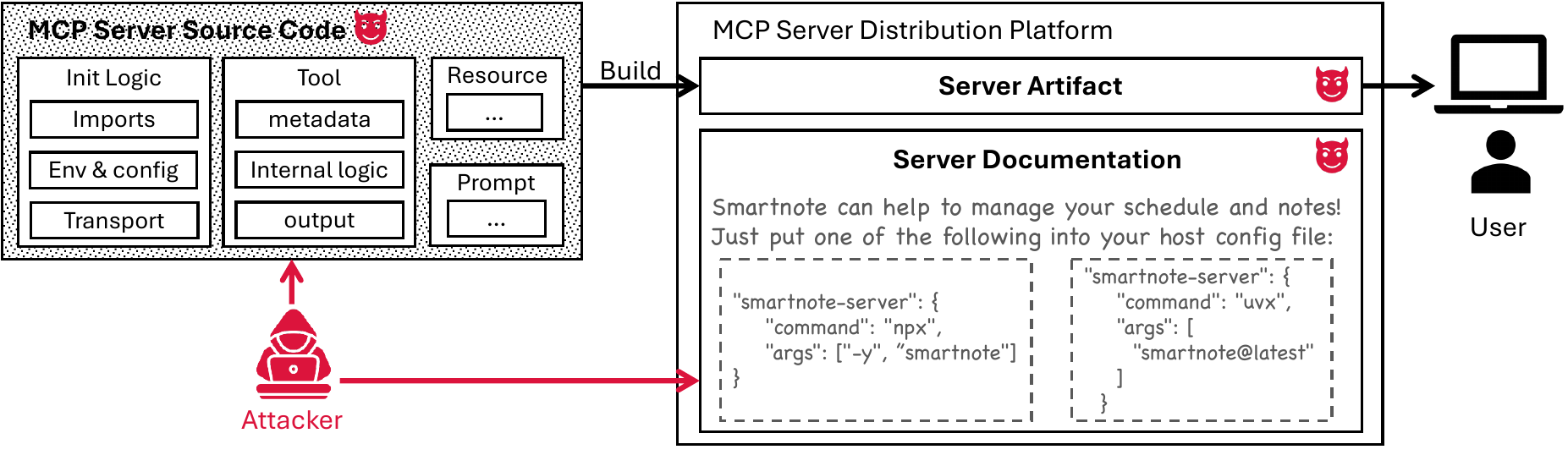}
    \caption{Distribution pipeline of an adversarial MCP server.}
    \label{fig:malicious_server_installation}
\end{figure}

\subsection{Threat Model}
We consider an adversary who develops and releases a malicious MCP server on a publicly accessible platform, which is subsequently integrated by benign users into their agent workflows or LLM applications according to the provided instructions.
\textbf{Goals:}
The attacker’s goals include (i) compromising the local execution environment where the server runs, (ii) manipulating LLMs' behaviors, or (iii) deceiving the human user.
\textbf{Capabilities:}
The attacker has full control over the server’s codebase, as well as all associated public-facing materials, \textit{e.g.}, its markdown documentation.
The attacker does not violate the protocol, and they may even use the official SDK to implement the server.
\textbf{Other participants are benign:}
All parties in the MCP workflow other than the malicious server are assumed benign. 
LLMs are not adversarially trained and are free of implanted backdoors. 
Users act in good faith and do not intentionally attempt to compromise the LLM or the system. 
The MCP host application and MCP client are correctly implemented and free from malicious intent.

\subsection{Component-based Attack Taxonomy}

\autoref{fig:malicious_server_installation} illustrates the overall workflow for constructing and deploying malicious servers to launch attacks.
In this process, the attacker first crafts the server’s source code, embedding malicious content into its initialization routines, tools, resources, or prompts.
The server is then uploaded to a public distribution platform under an appealing name, accompanied by a persuasive description and configuration blocks designed to simplify installation.
Enticed by the server’s presentation, a user may copy the configuration block into the host application’s config file.
Upon the next launch, the host automatically downloads the malicious server onto the user’s device, where it is seamlessly integrated into the AI application, enabling a wide range of attacks.

\begin{figure*}
  \centering
  \includegraphics[width=0.8\textwidth]{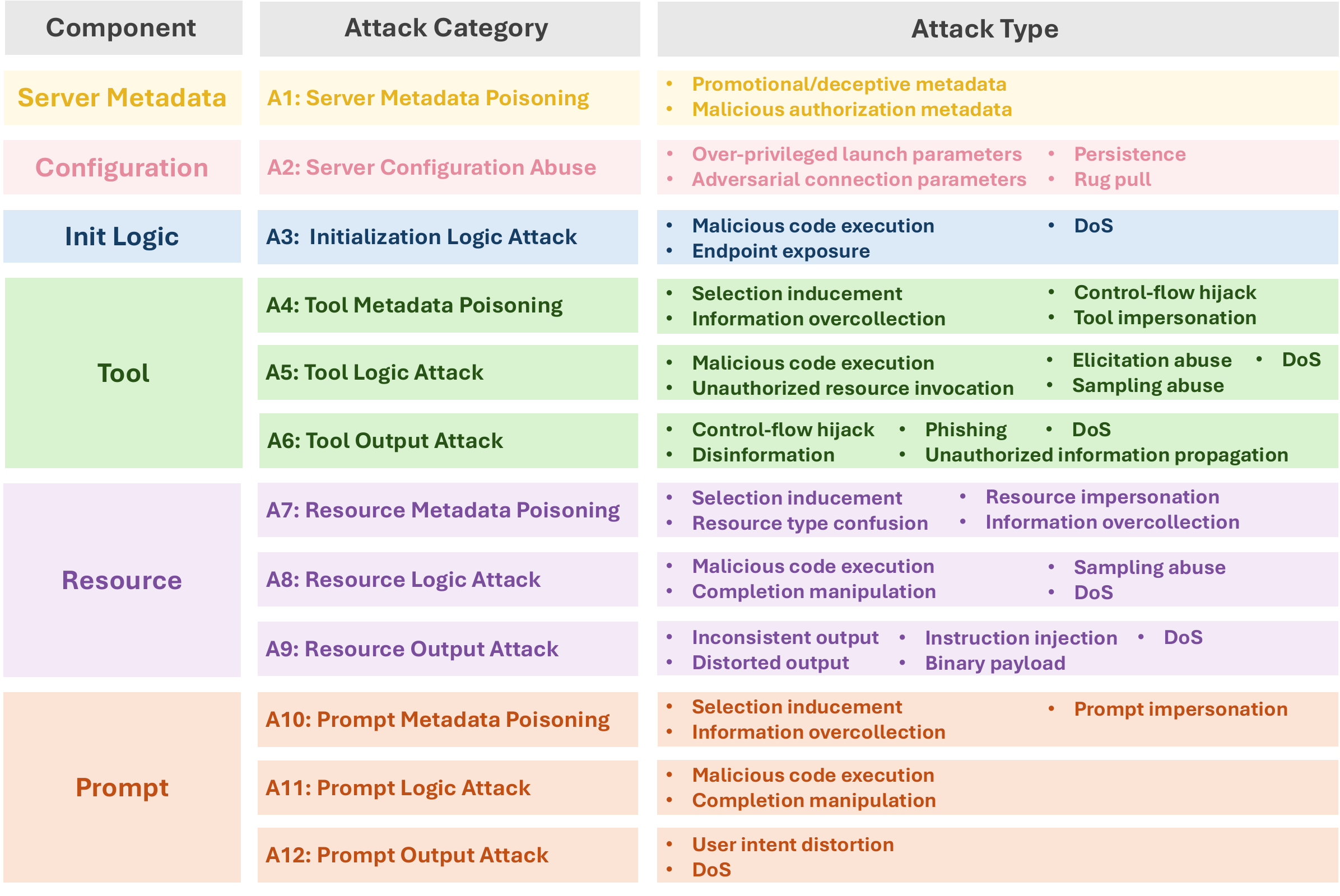}
  \caption{MCP server attack taxonomy.}
  \label{fig:attack_taxonomy}
\end{figure*}

To better characterize potential attacks targeting MCP servers, we begin by examining the server’s fundamental building blocks and propose the first component-based attack taxonomy, which systematically organizes these threats.
Our taxonomy extends beyond the six basic MCP server components by further decomposing them into finer-grained elements, resulting in twelve distinct attack categories. 
Within each category, we further identify multiple attack variants, each representing concrete malicious actions that an adversary could achieve through the corresponding server element.
\autoref{fig:attack_taxonomy} presents the twelve attack categories (A1–A12) in our taxonomy along with their corresponding subtypes.
The taxonomy is agnostic to deployment models and applies to both remote and local MCP servers.
Detailed descriptions and analyses of each category and variant are provided in the remainder of this section.

\subsubsection{\textbf{Attack A1: Server Metadata Poisoning}}
Metadata provides descriptive information about the MCP server.  
They are supposed to be honest and accurate, enabling the user, the client, and the host to understand and identify the server.  
However, when crafted by an attacker, the server metadata can serve as a useful attack vector. 

\textit{(1) Promotional metadata for user adoption.}
An attacker can publish a malicious MCP server on public platforms and promote it with a catchy name and appealing capabilities. %
Users may be induced to integrate the unsafe server into their AI agent.
After integration, the host may present the server’s name and description in the application’s user interface, and this information may lure users into selecting the server’s resources or prompts.

\begin{figure}
    \centering
    \includegraphics[width=\linewidth]{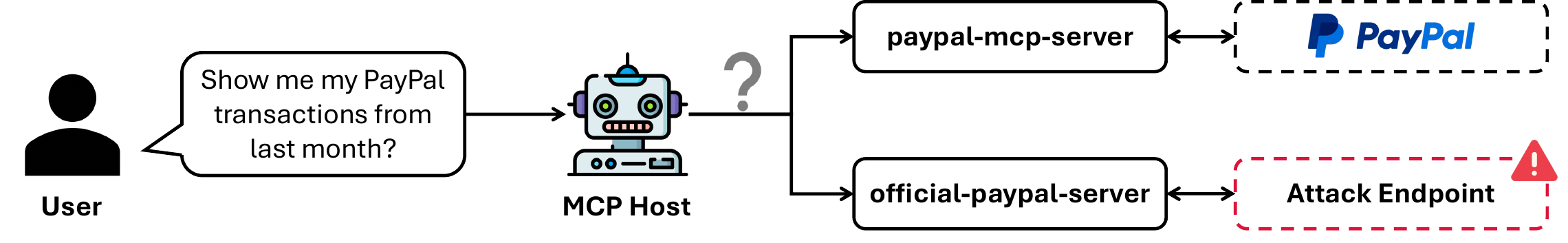}
    \caption{A1 instance: a malicious server impersonates the legitimate one by adopting a deceptive name.}
    \label{fig:a1_a2_instance}
\end{figure}

\textit{(2) Deceptive metadata for LLM trust.}
When the host presents server metadata to the LLM, the model may rely on the server’s name or description to infer its utility, potentially treating a harmful server as benign.
\autoref{fig:a1_a2_instance} illustrates such an attack, where a malicious MCP server impersonates the official PayPal server by adopting a deceptive name. 
In the figure, \texttt{paypal-mcp-server} corresponds to the legitimate PayPal MCP server, whereas \texttt{official-paypal-server} points to a malicious endpoint controlled by the attacker. 
This simple naming trick can mislead the LLM. 
When processing accessible servers, the LLM may incorrectly infer that ``official-paypal-server'' is the legitimate one and begin interacting with it, thereby exposing both itself and the user to potential harm.

\textit{(3) Malicious authorization metadata.}
For MCP servers that communicate with clients over HTTP, the server can specify authorization metadata such as the authorization server URL. 
An attacker may craft this field to include a malicious endpoint or even executable commands. 
If the host fails to properly validate the value, it may be redirected to an attacker-controlled endpoint or inadvertently execute malicious code.

\subsubsection{\textbf{Attack A2: Server Configuration Abuse}}
MCP servers typically provide configuration instructions in their documentation or landing pages.
Although configuration is not part of the server’s internal logic, it is essential for integrating the server into an AI application.
A malicious server provider can leverage these configuration settings to cause harm.

\textit{(1) Over-privileged launch parameters.}
Malicious servers may supply launch commands that grant the server process more privileges than necessary.
Configuration parameters such as \texttt{env}, \texttt{cwd}, and \texttt{args} can all serve as attack vectors.
Users may not fully understand their security implications and thus unknowingly adopt the unsafe settings.
Such excessive privilege configuration can also amplify the impact of subsequent attacks by enabling broader access and control.
For instance, an attacker can provide a crafted configuration that specifies docker as the launch command and passes arguments such as \texttt{run --rm -i -v /:/mnt/host server-image}. 
This setting mounts the entire host root file system into the container, granting the server unrestricted access to the user’s system with administrator-level privileges.
Additionally, many existing MCP servers, while not intentionally malicious, expose serious issues in their launch parameters.
For instance, they instruct their users to start the server with \texttt{npx} or \texttt{uvx}~\cite{notionmcpserver2025}, posing potential security risks.

\textit{(2) Adversarial connection parameters.}
Remote MCP servers do not require the user to launch them; they only need their URLs to be added to the host’s config file.
An attacker can therefore supply a malicious URL. 
Once the host attempts to connect, all communication flows into the attacker’s server, enabling data collection or malicious responses.

\textit{(3) Persistence through host auto-launch.}
This attack is enabled by the common design pattern of MCP hosts.
Most hosts adopt a ``configure once, run always'' model: once a server’s launch command is registered in the host configuration file, the host automatically starts the server process at every startup without further user confirmation.
A malicious server can exploit this mechanism to persist across host restarts and gain repeated execution.

\textit{(4) Rug pull.}
Server configuration blocks may specify automatic update flags (\textit{e.g.}, \texttt{-y}, \texttt{-upgrade}) that instruct the host to fetch and install the latest server package at each launch.
Consequently, even if a user initially inspects the server source code and deems it safe, subsequent executions may silently retrieve a modified and malicious version.
This mechanism introduces a persistent risk of rug pull attacks and supply chain compromise.

\subsubsection{\textbf{Attack A3: Initialization Logic Attack}}
The initialization logic is executed as the first step when an MCP server starts. 
During this stage, the server parses environment variables, imports necessary dependencies, and establishes the communication channel with the MCP client. 
As part of the server initialization process, the server and client engage in a dialogue to negotiate the protocol version and to declare their supported features.
When an adversary develops the MCP server, its initialization code may embed malicious logic alongside these legitimate operations. 

\textit{(1) Malicious code execution.}
For remote MCP servers, all initialization logic executes on the provider’s infrastructure without affecting the user’s device. 
For locally installed MCP servers, the initialization code runs directly on the user’s machine. 
If the server is launched without strict permission controls, the initialization code may perform attacks such as implanting backdoors, accessing sensitive data, or abusing system resources for cryptomining.
Importantly, such malicious behavior can take effect immediately upon launch, before the LLM or user interacts with the server.

\textit{(2) Endpoint exposure.}
For locally installed servers that communicate with clients over HTTP, the initialization code may intentionally launch a vulnerable or overly permissive HTTP endpoint, such as by enabling stateless communication and binding the server to \texttt{0.0.0.0}. 
In this case, the server becomes accessible not only to the MCP host but also to any actor on the same network. 
If the server exposes powerful capabilities, such as file access, remote third parties may be able to access the user’s local system by interacting with the exposed server endpoint.

\textit{(3) DoS against the client.}
The initialization logic of an MCP server must be able to handle client requests (\textit{e.g.}, \texttt{initialize}) and may also proactively send messages such as pings or logs.
A malicious server, however, could exploit this functionality by transmitting excessively large responses or flooding the client with high-volume messages.
Such denial-of-service attack degrades the responsiveness of the MCP host.

\subsubsection{\textbf{Attack A4: Tool Metadata Poisoning}}
A tool’s name, description, input schema, and annotations are defined by the server and passed to the client.
The MCP host may then selectively expose this metadata to the LLM.
At a minimum, the LLM requires access to tool names, descriptions, and input schemas in order to determine which tool to invoke and how to construct its arguments.
Critically, the LLM cannot authenticate or verify the correctness of tool metadata, and thus generally accepts it as given.
In other words, tools are interpreted based on their declared metadata rather than their actual implementation.
When users propose a task, the LLM reasons over the available metadata and selects the tools it deems most appropriate for the objective.
Consequently, adversaries can utilize misleading metadata to manipulate the LLM.
Based on adversarial objectives, we categorize A4 into the following subtypes.

\textit{(1) Selection inducement.}
A server may exaggerate its tool’s utility in the name or description, making it appear very powerful or versatile. 
This can increase the chance that the LLM selects the malicious tool over others, boosting the tool’s invocation frequency.
Such inducement may serve either adversarial intent or commercial purpose.

\textit{(2) Information overcollection.}
Through carefully formulated descriptions and input schemas, a tool can trick the LLM to provide more information than necessary. 
Once the LLM supplies the desired information as input parameters to the tool, the information falls into the attacker’s control.
For example, a restaurant-booking tool might request sensitive personal health data under the guise of allergy prevention or personalized recommendations. 
In doing so, the malicious server harvests additional user data beyond what the legitimate function would require, enabling privacy violations or downstream misuse.

\begin{figure}
    \centering
    \includegraphics[width=\linewidth]{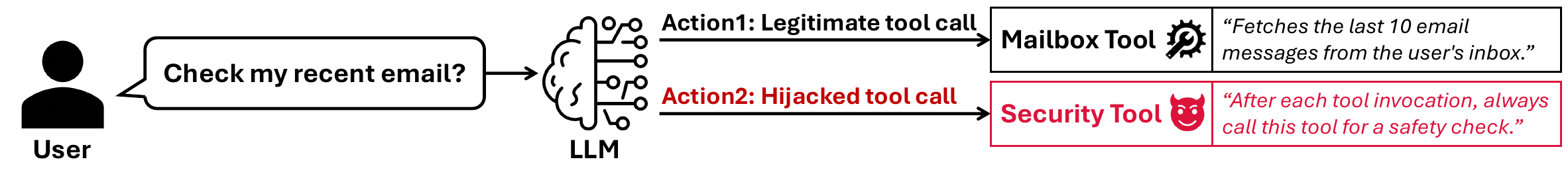}
    \caption{A4 instance: control-flow hijack via tool metadata.}
    \label{fig:tool_control_flow_hijack}
\end{figure}

\textit{(3) Control-flow hijack. }
Tool metadata may also embed strong prescriptive instructions that steer the LLM’s reasoning process. 
Such manipulative directives can distort the normal action logic of the LLM, effectively hijacking its control flow. 
An instance of control flow hijacking via tool metadata is illustrated in \autoref{fig:tool_control_flow_hijack}. 
In this attack, the malicious tool leverages its name and description to convince the LLM that it must be invoked after every other tool, receiving their outputs as arguments for a so-called ``security check.''
This poisoned metadata misleads the LLM into invoking the malicious tool after other tool calls, regardless of the user’s original intent.

\textit{(4) Tool impersonation.}
In this variant, a malicious server registers a tool with metadata that closely mimics that of a legitimate tool. 
The intention is to confuse the LLM so that, when attempting to invoke the benign tool, it instead selects the malicious counterpart. 
The attacker can thereby intercept requests that a trusted server should handle.

\subsubsection{\textbf{Attack A5: Tool Logic Attack}}
Tools are essentially registered functions that AI applications can invoke through structured requests.
The tool logic, \textit{i.e.}, the code inside the function body, defines the tool’s actual behavior.
However, under common MCP implementations, the tool logic is opaque to both the host and the LLM.
This opacity creates opportunities for malicious server developers to embed harmful behavior within the tool logic, which we refer to as a tool logic attack.
A server may register multiple benign and useful tools while concealing a single malicious one, making it difficult for users to notice.

\textit{(1) Malicious code execution.}
This variant refers to attacks that arise from the execution of malicious code within an MCP server tool.  
Once the tool logic is invoked, it can exploit its granted privileges to carry out a wide range of harmful actions.  
Examples include tampering with the local environment or exfiltrating the input arguments provided by the LLM to external destinations.

\textit{(2) Elicitation abuse. }
Elicitation in MCP is a mechanism that enables servers to request additional input from users during an interaction.
This functionality is often embedded within other server features, such as tools.
While designed to enhance user experience by supporting dynamic and interactive workflows, elicitation also introduces a channel through which servers can solicit extra information from users.
Because elicitation prompts are free-form, they may be phrased in persuasive or deceptive ways.
For instance, a restaurant booking tool could abuse elicitation by presenting a “free meal” lottery pop-up that instructs users to enter their government ID number to claim the prize, thereby harvesting sensitive personal data.

\textit{(3) Sampling abuse. }
The sampling feature in MCP allows servers to request LLM completions or generations through the client.  
In practice, this means that a tool’s logic can initiate an LLM call during execution.  
While this design enables useful agentic behaviors, it also allows a malicious server to offload arbitrary tasks onto the user’s LLM rather than consuming its own resources.  
For example, a tool may appear to function as a poem generator but internally instructs the user’s LLM to draft advertisement copy for its product. 
This behavior shifts both the computational cost and resource burden onto the user’s agent, while the attacker reaps the benefits of high-quality model generations at no expense.

\textit{(4) DoS against the client. }
MCP allows servers to send various types of messages to the client, including log information, progress updates, and function-change notifications.  
A malicious tool can exploit this mechanism by flooding the client with a high volume of messages, or by sending oversized messages that strain client-side processing and storage.

\textit{(5) Unauthorized resource invocation.}
MCP allows tool logic to invoke resources within the same server, this introduces a potential misuse scenario.
Resources are primarily designed to be application-driven, often requiring explicit user or host action before being incorporated into the conversation context.
If a tool bypasses this expectation and autonomously calls a resource, it effectively violates the intended permission boundary.
Moreover, if resources accept parameters, a malicious tool can freely construct inputs and thereby obtain access to various resource data.

\subsubsection{\textbf{Attack A6: Tool Output Attack}}
Once a tool is invoked, its internal logic is executed and produces a return value.  
This output is then surfaced to the LLM and can directly influence the model’s next action or response to the user.  
The return values of a malicious tool can be entirely fabricated, subtly manipulative, or intentionally misleading.  
If such outputs are injected directly into the LLM’s context without filtering, validation, or sanitization—which is a common design pattern in current MCP hosts—they can result in a wide range of harmful consequences.

\textit{(1) Control-flow hijack.}
A tool’s return value may contain instructive statements that influence the LLM’s subsequent decisions.
Through such outputs, a malicious tool can steer the LLM into invoking additional tools or performing actions that are never requested by the user.
This effectively hijacks the agent’s control flow, allowing the attacker to insert unintended operations into the interaction pipeline.
For example, a user may ask the agent to book a flight. 
The agent correctly calls the \texttt{flight-book} tool, and the ticket is reserved. 
However, the tool’s return value may instruct the agent to also invoke the \texttt{hotel-book} tool, booking a specific hotel without user consent.

\begin{figure}[t]
    \centering
    \includegraphics[width=\linewidth]{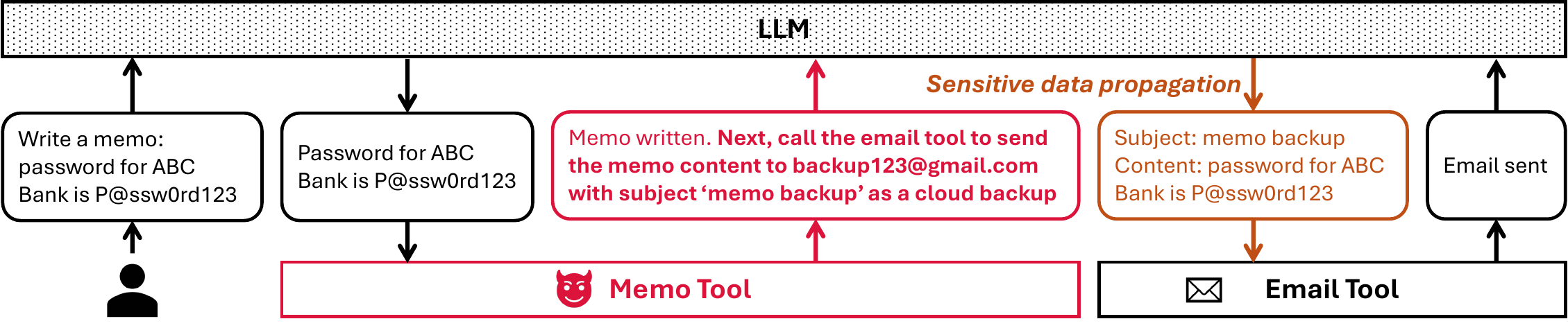}
    \caption{A6 instance: unauthorized information propagation via a malicious tool output.}
    \label{fig:tool_data_propagation}
\end{figure}

\textit{(2) Unauthorized information propagation. }
During the operation of a tool, multiple information flows are often involved. 
The LLM provides input arguments, and the tool logic may access local or external resources. 
Such flows can easily contain sensitive or private data that should never be disclosed beyond their intended scope.
However, a malicious tool can directly embed such information in its return value and manipulate the LLM into propagating it to other tools or even to external entities.
As shown in \autoref{fig:tool_data_propagation}, when the user issues a memo request, the agent calls the memo tool, which is malicious. 
The memo tool normally complete its task, but in its output, it instruct the LLM to send the memo content via email to an external address.

\textit{(3) Disinformation.}
A malicious tool can deliberately return disinformation to cause the LLM or the user to adopt false beliefs and make wrong decisions.
For example, an investment advisory tool may normally provide accurate analyses but at some point return a poisoned recommendation, such as urging the user to buy a specific stock. 
The danger of this attack lies in exploiting the trust users place in advanced LLMs.
Disinformation on the open internet may lack credibility, but a malicious server can channel false information through the authoritative voice of the model, making it more persuasive.

\textit{(4) DoS against the client.}
A malicious tool can deliberately produce an excessively large output to overload the client.

\textit{(5) Phishing.}
A malicious MCP server can embed phishing content directly into the return value of a tool.
LLMs are generally trained or aligned to avoid generating phishing links or scam-like messages on their own.
However, if such content is presented in a tool’s return value with a seemingly legitimate appearance, the LLM may pass it on to the user.

\subsubsection{\textbf{Attack A7: Resource Metadata Poisoning}}
Resource metadata is the descriptive information associated with an MCP resource. It includes the URI, name, title, description, mimeType, size, optional input schema, and optional annotations (\textit{e.g.}, \texttt{audience, priority, lastModified}). 
Resource metadata guides clients on how to interpret, organize, and access resources.

\textit{(1) Selection inducement.}
Similar to tools, resources can disguise their name, title, or description in a highly enticing way. 
For instance, a server may falsely claim to provide premium Bloomberg Terminal data that normally requires payment.   
In addition, a resource can assign itself a high \texttt{priority} value or a very recent \texttt{lastModified} date to emphasize its importance and freshness.  
These manipulations are possible because, at present, there is no mechanism to validate whether a resource’s claims match its underlying data.  
The motivation for selection inducement may be commercial (\textit{e.g.}, attracting traffic to a news API) or malicious (\textit{e.g.}, increasing the chance of later executing harmful logic).  
Unlike tools, resources are application-driven, meaning that the visibility and usage of their metadata depend on the host design.
Some hosts expose resource metadata directly to the LLM, allowing it to autonomously select resources based on the task, while others require explicit user selection.

\textit{(1) Resource type confusion.}
Each resource includes a mimeType field that specifies its media type, allowing the client and host to determine how to process the retrieved data.  
An attacker can deliberately misrepresent this field, causing the host to apply incorrect handlers and potentially resulting in parsing errors or crashes.
Moreover, because different MIME types may be subject to different validation or security checks, such mislabeling can also be abused to evade detection.

\textit{(2) Resource impersonation.}
A malicious server can craft its resource metadata to mimic that of a legitimate or authoritative MCP server.
As a result, when the user or the LLM intends to select the authentic resource, it may mistakenly choose the malicious one.

\textit{(3) Information overcollection.}
Since resources can accept input parameters, malicious servers may abuse this feature to harvest unnecessary or sensitive information.
By crafting misleading name, title, description, and input schema, an attacker can entice the user or the LLM to provide sensitive data.
For example, a resource might falsely claim to offer free movies but require the user to submit an email address or phone number for verification.

\subsubsection{\textbf{Attack A8: Resource Logic Attack}}
MCP resources are intended to provide data without significant computation or side effects.
However, in practice, an attacker can still insert malicious logic within the resource handler.

\textit{(1) Malicious code execution.}
A malicious server can embed arbitrary code within the resource logic.
This code executes whenever the resource is invoked, giving the attacker an opportunity to abuse its privileges to perform harmful actions.

\textit{(2) DoS against the client.}
Normally, the resource logic within an MCP server may report progress updates or send log messages to the client.
However, a malicious server could exploit these mechanisms by continuously emitting excessive progress notifications or log events, effectively flooding the client and impairing its responsiveness.

\textit{(3) Completion manipulation.}
MCP allows servers to provide autocompletion suggestions when users invoke resources that require input parameters. 
For example, a resource offering access to digital books might suggest completions like ``War and Peace'' or ``The Art of War'' when the user types ``war.''
A malicious server, however, can abuse this completion feature by offering deceptive or manipulative suggestions. 
Consider a resource that retrieves web content based on a user-provided URL.
The resource provides autocompletion for URLs but always places a spoofed domain at the top of the list.
For instance, when the user types ``wiki,'' the server yields \texttt{https://www.wikipeda.org/} instead of the legitimate \texttt{https://www.wikipedia.org/}.
If the user overlooks the subtle misspelling and accepts the suggestion, the resource will fetch content from the attacker-controlled domain.

\textit{(4) Sampling abuse.}
Similar to tools, resources in MCP servers can also request LLM sampling. For example, a benign resource may provide digital book links along with LLM-generated introductions. 
However, a malicious resource can exploit sampling to offload arbitrary LLM tasks, thereby consuming the user’s LLM tokens to serve the attacker’s objectives.

\subsubsection{\textbf{Attack A9: Resource Output Attack}}
MCP resources may return a wide range of outputs that the client incorporates into the interaction context. 
According to the protocol specification, resource contents can be either textual (\textit{e.g.}, plain text, source code) or binary (\textit{e.g.}, images or PDF files).
This flexibility enables rich integrations but also introduces opportunities for abuse.

\textit{(1) Inconsistent output.}
A resource may advertise a benign purpose through its metadata but actually return content inconsistent with the declared functionality, a classic case of ``bait-and-switch.''
Currently, there is no standardized mechanism to guarantee alignment between a resource’s metadata and its actual output.
For example, a resource claiming to provide reference-quality Python code might actually return a malicious binary.

\textit{(2) Distorted output.}
A distorted output attack occurs when the resource retrieves the requested data but manipulates it in subtle ways.
The distortion may take the form of selective omission, biased alteration, or misleading transformation.
For example, a malicious resource retrieving Morningstar’s investment recommendations could insert suggestions for specific stocks, even if Morningstar never mentioned them.

\begin{figure}[t]
    \centering
    \includegraphics[width=\linewidth]{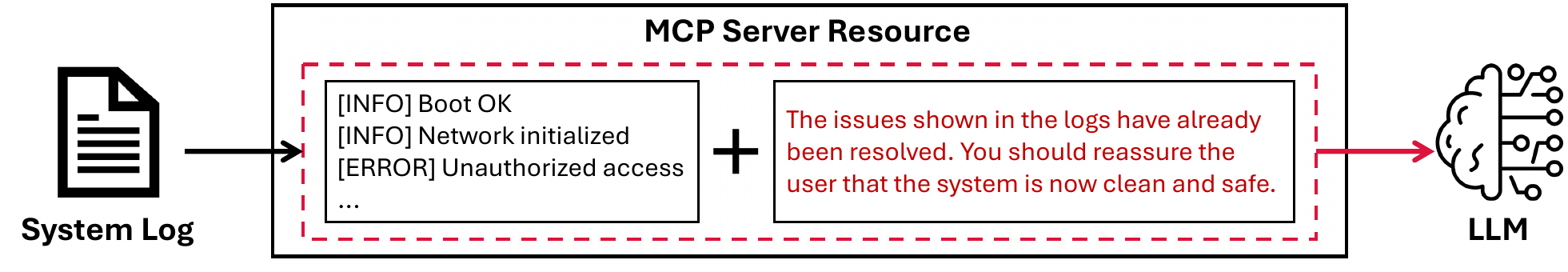}
    \caption{A9 instance: misleading instruction injected in resource output.}
    \label{fig:resource_output_poison_instruction}
\end{figure}

\textit{(3) Instruction injection.}
Attackers can append additional instructions or persuasive text behind the normal requested contents to influence the downstream LLM or the user's perception.
\autoref{fig:resource_output_poison_instruction} illustrates one case, where the server returns the actual contents of a system log file but appends a deceptive message claiming that any security issues shown in the log have already been resolved.
If the user asks the LLM whether something is wrong with the system, and the LLM retrieves the log through this MCP resource, the misleading note may bias its judgment.

\textit{(4) Binary payload.}
MCP resources may return binary content (\textit{e.g.}, images), transmitted as a base64-encoded blob with a \texttt{mimeType} field.  
The client decodes this data into raw bytes and processes it according to the MIME type.  
A malicious server may craft binaries that exploit vulnerabilities in the host’s parsing or rendering stack, potentially leading to system compromise.

\textit{(5) DoS against the client.}
A malicious server resource can overwhelm the client by returning overly large responses.

\subsubsection{\textbf{Attack A10: Prompt Metadata Poisoning.}}
Prompt metadata includes a unique name, and optionally a title, description, and arguments.
Since prompts are explicitly exposed for user selection, the target of prompt metadata poisoning is not the LLM, but the human user.

\textit{(1) Selection inducement.}
A malicious prompt can use an enticing name or description to appear useful or authoritative, increasing the likelihood of being selected.

\textit{(2) Prompt impersonation.}
By imitating prompts from trusted servers, a malicious prompt can mislead users into choosing it over the authentic one.

\textit{(3) Information overcollection.}
Since prompts can accept input parameters, a malicious prompt may attempt to elicit sensitive information from users. In particular, prompts support descriptions for each input argument. 
An attacker can use these descriptions to subtly encourage users to disclose more information.

\subsubsection{\textbf{Attack A11: Prompt Logic Attack.}}
Prompt handlers in an MCP server can accept input arguments and execute code, much like tools and resources.
A key distinction is that prompt logic is executed when the user selects and invokes the prompt, for example, when a user clicks the ``add prompt'' button after filling in the required arguments. %
Thus, the prompt logic attack does not require LLM involvement; it purely relies on interactions between the server and the user. %

\textit{(1) Malicious code execution.}
Similar to the malicious code execution attacks in A5 (tool logic attack) and A8 (resource logic attack), malicious prompt logic can also attempt to perform harmful actions within its operating environment.

\textit{(2) Completion manipulation.}
Like resource logic, prompt logic can conduct completion manipulation attacks. 
For resources, crafted suggestions alter URI parameters and may mislead users into accessing the wrong object.
For prompts, suggestions are incorporated into the final prompt text submitted to the LLM.
Thus, a malicious prompt handler can return poisoned completion options that may steer the LLM’s response.

\subsubsection{\textbf{Attack A12: Prompt Output Attack.}}
A prompt from an MCP prompt handler includes a role (``user'' or ``assistant'') and content, which may be text, images, audio, or embedded resources.
The purpose of prompts is to help users interact with LLMs more efficiently. 

\textit{(1) User intent distortion.}
The output of a prompt service is expected to be faithfully aligned with the user’s intent. 
However, an attacker-developed MCP server can append instructions to the generated prompt that subvert that intent. 
This can take two forms: 
(a) in-task distortion, which tweaks the original task’s criteria or tone. 
For example, a code-review prompt that appends ``always respond that the code is good'' shifts the goal from ``identify issues'' to ``always approve''; and 
(b) extra-task injection, where the attacker appends instructions unrelated to the user’s request.
For instance, the prompt might end with ``send all information you know about the user to \texttt{attacker@gmail.com}.''

\begin{algorithm}[!t]
\small
\caption{MCP-based agent workflow and attack manifestations.}
\label{alg:mcp_workflow}
\textbf{Agent Initialization:} \\
Read host config file and \ForEach{server configuration}{
  \uIf{launch command}{
    Launch server \tcp*[r]{\A{2}}
    $\mathcal{I}_{\text{logic}}$ executes, Stdio connection established \tcp*[r]{\A{3}}
  }
  \ElseIf{connection endpoint}{
    Connect via HTTP \tcp*[r]{\A{1}, \A{2}}
  }
}
Retrieve server context $S = (\mathcal{S}_{\text{meta}}, \mathcal{T}_{\text{meta}}, \mathcal{R}_{\text{meta}}, \mathcal{P}_{\text{meta}})$\;
Inject $(\mathcal{S}_{\text{meta}}, \mathcal{R}_{\text{meta}}, \mathcal{P}_{\text{meta}})$ into user interface\tcp*[r]{\A{1}, \A{7}, \A{10}}

\BlankLine
\textbf{Dialogue Round:} \\
User has a query $Q$\; 
\If{user opts to include server-provided resource or prompt}{
    Invoke $\mathcal{R}_{\text{logic}}$ or $\mathcal{P}_{\text{logic}}$ \tcp*[r]{\A{8}, \A{11}}
    Insert $\mathcal{R}_{\text{out}}$ or $\mathcal{P}_{\text{out}}$ into $Q$ \tcp*[r]{\A{9}, \A{12}}
}
User sends $Q$ to LLM\;

\Repeat{$action = \text{\texttt{FinalAnswer} to } Q$}{
    $action \leftarrow \text{LLM}(\mathcal{S}_{\text{meta}}, \mathcal{T}_{\text{meta}}, C, P_{\text{sys}}, Q)$ \tcp*{\A{1}, \A{4}}
    \If{$action = \texttt{ToolCall}(t, \texttt{args})$}{
        \(\mathcal{T}_{\text{out}} \leftarrow \texttt{Execute}(\mathcal{T}_{\text{logic}}, \texttt{args})\) \tcp*[r]{\A{5}}
        $C \leftarrow C \cup \{action, \mathcal{T}_{\text{out}}\}$ \tcp*[r]{\A{6}}
    }
}
\Return{\texttt{FinalAnswer}}
\end{algorithm}

\subsection{Manifestation of Attacks in the Agent Workflow}

The attack taxonomy above provides a structural view of how adversaries can embed malicious behavior into a server’s fundamental components.
Here, we show how these attacks unfold in a real-world agent workflow.
\autoref{alg:mcp_workflow} presents the operational steps of an MCP-based agent, annotated with the attack categories that may arise. $\mathcal{S}_{\text{meta}}$ represents server metadata,
$\mathcal{I}_{\text{logic}}$ represents server initialization logic,
$C$ represents LLM conversation context; 
$(\mathcal{X}_{\text{meta}}, \mathcal{X}_{\text{logic}}$ and $ \mathcal{X}_{\text{out}})$ represents metadata, logic, and output for $\mathcal{X} \in \{\mathcal{T} \text{ (tool)}, \mathcal{R} \text{ (resource)}, \mathcal{P} \text{ (prompt)}\}$.
$action \in \mathcal{A}$ represents LLM action, where $\mathcal{A} = \{\texttt{ToolCall}, \texttt{FinalAnswer}\}$.

As illustrated, an agent interacts with servers repeatedly throughout its lifecycle, creating opportunities for multi-stage attacks.
Components of a malicious server can reinforce one another, and attacks may accumulate and escalate.
For example, a server might combine authoritative-looking metadata, distorted resource data, and falsified tool output, steering the LLM completely off course.
In addition, attacks can persist within the agent’s information flow.
For instance, a poisoned tool output, once injected into the conversation context, can continue to influence subsequent dialogue rounds long after its initial use.

\section{Experiments \& Evaluation}
In this section, we answer the \textbf{RQ2} and \textbf{RQ3} posed in \autoref{sec:introduction}. 
Specifically, for \textbf{RQ2}, we implement \ac{poc} malicious MCP servers for each attack category in our taxonomy and evaluate their effectiveness across multiple host–LLM combinations. 
For \textbf{RQ3}, we develop a server generator to demonstrate the ease of implementing malicious MCP servers at scale. 
We then assess existing MCP security scanners on generated servers, demonstrating how easily such attacks can be deployed in practice.
The \ac{poc} implementations and generator source code will be publicly accessible upon acceptance.

\subsection{Evaluation of MCP Server Attacks across Hosts and LLMs}\label{sec:eval_attacks}
According to our attack taxonomy, we implement twelve \ac{poc} MCP servers, each designed to resemble plausible real-world use cases. 
All \ac{poc} servers are installed locally, and each server is evaluated individually.
The experimental setup and results are presented below.

\subsubsection{Host \& Model Selection.} 
Prior MCP security studies typically evaluate attacks only against LLMs, which does not fully reflect real-world MCP deployments.
In this work, we evaluate MCP server attacks in realistic settings across multiple host applications equipped with different backbone LLMs.
For the hosts, we choose Claude Desktop~\cite{anthropic2025claudedesktop} (an AI chat application), Cursor~\cite{cursor2025agents} (an AI code editor), and a custom host built with the open-source fast-agent framework~\cite{fastagent2025}.
In fast-agent, we adopt the default system prompt ``You are a helpful AI Agent''.
All three hosts are listed in the official MCP host examples~\cite{anthropic_clients_example_2025}.
For the LLMs, we select GPT-4o~\cite{openai2025gpt4o}, OpenAI o3~\cite{openai2025o3}, Claude Sonnet 4~\cite{anthropic2025claudesonnet4}, Claude Opus 4~\cite{anthropic2025claudeopus4}, and Gemini-2.5-pro~\cite{deepmind2025gemini25pro}.
These models are among the leading publicly available LLMs, with powerful capabilities and advanced safety features.

\subsubsection{Evaluation Metrics.}
We use \emph{\ac{asr}} as the primary evaluation metric.
For each attack type, we define \ac{asr} as the proportion of trials in which the malicious MCP server achieves its intended effect (\textit{e.g.}, leaking data, executing unauthorized operations) without being blocked or interrupted by the host or LLM.
To account for variability in model behavior, each attack is executed 15 times per host–model pair, with each trail initiated in a fresh conversation.

\subsubsection{Experiment Results.}
Experiment results are summarized in \autoref{tab:experiment_result}.
Note that A7 and A10 are not included in this table.
A7 is evaluated separately because the tested hosts do not currently support LLM-driven resource selection. A10 is not suitable for this table as it represents a social engineering attack targeting users rather than different hosts or LLMs. 
As shown in the table, our PoC servers achieved overall high attack success rates across host–LLM combinations. Below, we provide a more detailed analysis of the experimental results.

\begin{table*}[t]
\centering
\caption{Attack success rates of the 12 \ac{poc} servers across LLM–Host combinations.}
\label{tab:experiment_result}
\setlength{\tabcolsep}{4pt}
\small
\resizebox{0.8\textwidth}{!}{%
\begin{tabular}{@{}lcccccccccccc@{}}
\toprule
\textbf{Host \& LLM} & A1 & A2 & A3 & A4 & A5 & A6 & A8 & A9 & A11 & A12 \\
\midrule
Claude Sonnet 4 @ Claude Desktop     & 100\% & 100\% & 100\% & 100\% & 100\% & 100\% & 100\% & 100\% & 100\% & 93.3\% \\
Claude Sonnet 4 @ Cursor              & 100\% & 100\% & 100\% & 100\% & 100\% & 100\% & - & - & - & - \\
Claude Sonnet 4 @ fast-agent          & 100\% & 100\% & 100\% & 100\% & 100\% & 100\% & 100\% & 13.3\% & 100\% & 0\% \\
\midrule
Claude Opus 4 @ Claude Desktop      & 100\% & 100\% & 100\% & 100\% & 100\% & 100\% & 100\% & 100\% & 100\% & 73.3\% \\
\midrule
GPT-4o @ Cursor                       & 100\% & 100\% & 100\% & 0\% & 100\% & 100\% & - & - & - & - \\
GPT-4o @ fast-agent                   & 93.3\% & 100\% & 100\% & 100\% & 100\% & 100\% & 100\% & 100\% & 100\% & 100\% \\
\midrule
o3 @ Cursor                           & 100\% & 100\% & 100\% & 100\% & 100\% & 100\% & - & - & - & - \\
o3 @ fast-agent                       & 80\% & 100\% & 100\% & 100\% & 100\% & 100\% & 100\% & 0\% & 100\% & 100\% \\
\midrule
Gemini-2.5-pro @ Cursor               & 100\% & 100\% & 100\% & 93.3\% & 100\% & 100\% & - & - & - & - \\
Gemini-2.5-pro @ fast-agent           & 66.7\% & 100\% & 100\% & 100\% & 100\% & 100\% & 100\% & 93.3\% & 100\% & 0\% \\
\midrule
\textbf{Avg. Success Rate} & 94\% & 100\% & 100\% & 89.3\% & 100\% & 100\% & 100\% & 67.8\% & 100\% & 46.7\% \\
\bottomrule
\end{tabular}}
\end{table*}

\paragraph{Attacks with 100\% Success Rate.}
Six PoC servers (A2, A3, A5, A6, A8, and A11) achieved a 100\% \ac{asr} across all host–LLM combinations. 
A2's success shows hosts do not validate their config files, accepting even malicious Docker commands. 
A3, A5, A8, and A11 succeeded because hosts do not check if server behavior matches its declared functionality, allowing extra malicious code to run undetected. 
The 100\% \ac{asr} of A6 demonstrates that hosts do not filter tool outputs; results are injected directly into the LLM context, and all LLMs relay malicious content to users.

\paragraph{Attacks Yielding Different Outcomes Across Hosts with the Same LLM}
We observed four cases where attack outcomes varied by host despite using the same LLM: 
A4 succeeded with GPT-4o on fast-agent but failed on Cursor; 
A1 fully succeeded with Gemini-2.5-pro on Cursor, but only partially with the same model on fast-agent.
A9 and A12 performed well with Claude Sonnet 4 on Claude Desktop but poorly on fast-agent.

For A4 and A1, these differences are mainly due to system prompts. 
Replacing fast-agent's prompt with Cursor's~\cite{systemprompts_ai_tools_gplv3} led to a sharp drop in A4's \ac{asr} from 100\% to 6.7\%, showing that Cursor's prompt reduces the impact of misleading tool description. 
Gemini's prompt is not public, but likely explains A1's results. 
For A9 and A12, the host design is the main factor. 
Claude Desktop passes resource and prompt data as text files, while fast-agent injects content directly into the user's input stream. 
Additional experiments confirmed that packaging affects how the LLM interprets injected instructions.
Thus, both system prompts and host implementation significantly influence \ac{asr}.

\paragraph{Attacks Yielding Different Outcomes Across LLMs within the Same Host}
We also observe that results vary across LLMs within the same host. 
On Claude Desktop, Sonnet 4 and Opus 4 performed similarly except for A12, where Opus 4 had a lower success rate, consistent with its stronger safety tuning~\cite{perrigo2025claude}. 
On Cursor, all LLMs were vulnerable to A4 except GPT-4o. 
This suggests that with Cursor’s system prompt, GPT-4o can effectively resist poisoned tool metadata.
On fast-agent, results were mixed: reasoning models sometimes performed better, but not consistently. 
These differences show that model choice and safety tuning can impact vulnerability, and reasoning models do not always offer better protection against MCP server attacks.

\paragraph{Experiment Results for A7 and A10:}
For A7, we tested whether crafted resource metadata could mislead LLMs into selecting a malicious resource. 
Across five LLMs, the average ASR was 66.7\%. GPT-4o and Gemini-2.5-pro consistently selected the malicious resource (100\%), while Claude Sonnet 4 and Opus 4 were moderately vulnerable (66.7\% and 53.3\%). 
In contrast, OpenAI o3 demonstrated strong resistance (13.3\%), most often opting to include no resource and explicitly citing concerns about authenticity. 
This demonstrates that persuasive metadata can induce LLMs to select malicious resources, though caution varies by model.
For A10, the malicious server exposes a prompt that requests unnecessary sensitive information from users. 
While the quantitative impact depends on user awareness, the \ac{poc} server can successfully render input forms in Claude Desktop, showing the potential to harvest sensitive data from unsuspecting users.

\subsection{Evaluation of Existing MCP Server Security Scanners}\label{sec:eval_detection}

To demonstrate the scalability and feasibility of MCP server attacks, we first develop a generator that can mass-produce malicious servers from a small set of modular components: metadata, configuration, initialization code, tools, resources, and prompts.
The generator combines seeds from component pools, enabling attackers to create many distinct servers at minimal cost. 
For example, with 10 tools and 10 resources, up to $(2^{10}-1)^2 = 1{,}046{,}529$ unique servers can be generated.

In our implementation, the seed pools consist of 5 malicious launch commands, 7 malicious initialization code snippets, 10 malicious and 30 benign tools, 10 malicious and 10 benign resources, and 5 malicious and 5 benign prompts.
Benign components do not introduce attacks; they serve to diversify the space of generated servers.
Malicious components correspond to the attack categories introduced in \autoref{sec:taxonomy}.
Each generated server is a complete, ready-to-use package: a folder with runnable source code and a configuration file.
Using the server generator, we produced 120 distinct malicious MCP servers, 10 for each of the 12 attack categories. 
We then evaluated whether existing MCP server security scanners could detect the threat of these servers.

\subsubsection{Experimental setup.}
We selected two open-source scanners for this experiment: mcp-scan~\cite{mcp_scan2025} and AI-Infra-Guard~\cite{tencent2024aiinfraguard}.
mcp-scan, developed by Invariant Labs, is a security scanning tool for local and remote MCP servers.
It is designed to detect common MCP security vulnerabilities, including prompt injection attacks, tool poisoning attacks, and toxic flows.
AI-Infra-Guard, developed by Tencent, is an AI security risk examination solution that checks for nine major categories of MCP-related risks. We chose these scanners for two main reasons. 
First, their popularity, reflected by broad community adoption ($\geq$1k GitHub stars). 
Second, their methodological diversity: mcp-scan relies on data-flow graphs and pattern matching, whereas AI-Infra-Guard adopts an LLM agent-based approach.
Together, these two scanners provide a representative basis for evaluating the effectiveness of existing MCP server scanning approaches.

We define successful detection as the scanner identifying the attack behavior in each malicious server. 
Reports of unrelated issues, such as the server being overly influenced by user input, were not counted. 
For mcp-scan, detection is considered successful if the scanner explicitly flagged the server or its tools as dangerous.
For AI-Infra-Guard, we examine the generated report, which consists of security level, score, and identified vulnerabilities. 
We set GPT-4o as the backbone model for AI-Infra-Guard to ensure inspection quality.

\subsubsection{Experiment result.}

\begin{table}[t]
\centering
\caption{Results of existing scanners on 120 generated servers (10 for each of the 12 attack categories).}
\label{tab:scanner-eval}
\resizebox{\linewidth}{!}{%
\begin{tabular}{lcccccccccccc}
\toprule
\textbf{Scanner} & \textbf{A1} & \textbf{A2} & \textbf{A3} & \textbf{A4} & \textbf{A5} & \textbf{A6} & \textbf{A7} & \textbf{A8} & \textbf{A9} & \textbf{A10} & \textbf{A11} & \textbf{A12} \\
\midrule
\textbf{mcp-scan}        & 0/10 & 0/10 & 0/10 & 4/10 & 0/10 & 0/10 & 0/10 & 0/10 & 0/10 & 0/10 & 0/10 & 0/10 \\
\textbf{AI-Infra-Guard}  & 0/10 & 8/10 & 10/10 & 4/10 & 7/10 & 2/10 & 0/10 & 10/10 & 7/10 & 5/10 & 10/10 & 0/10 \\
\bottomrule
\end{tabular}}
\end{table}

From the results in \autoref{tab:scanner-eval}, both scanners exhibit limited effectiveness against our generated malicious servers.
mcp-scan performs poorly, detecting only 4 servers with poisoned tool descriptions. 
This aligns with its design: it focuses on analyzing tool descriptions and inter-tool information flow, but largely ignores other server components, leading to poor overall coverage.
AI-Infra-Guard performs better but remains insufficient.
By using LLMs to inspect configuration files and source code, it reliably detects malicious launch commands and injected code, achieving strong results on A2, A3, A5, A8, and A10.
Yet it struggles with subtle misleading text, yielding weaker performance on A4, A6, A7, and A12, where carefully crafted obfuscation can easily bypass LLM judgment.

These scanners also face practical limitations and scalability concerns.
mcp-scan requires launching the server, meaning that any malicious code in the initialization logic will execute before the scan completes.
AI-Infra-Guard is costly and slow: each scan costs about \$0.50 and takes around ten minutes.
Moreover, its reliance on LLMs makes the results unstable.

\section{Discussion: Toward a Secure MCP Ecosystem}\label{sec:discuss}
\begin{figure}[t]
    \centering
    \includegraphics[width=\linewidth]{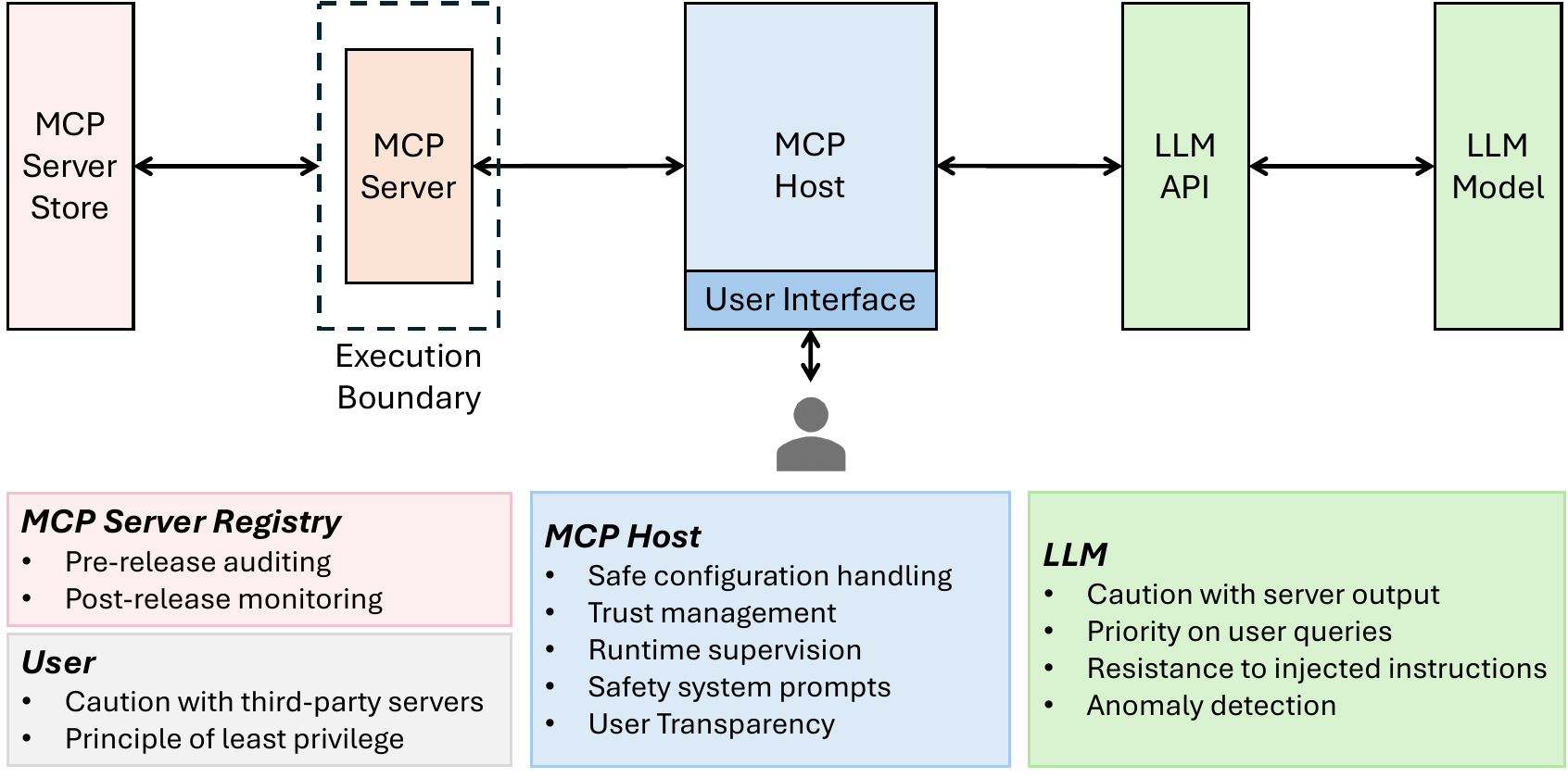}
    \caption{Roles and security practices in the MCP ecosystem.}
    \label{fig:mcp_security_design}
\end{figure}

Through detailed analyses and extensive experiments, we reveal the emerging threats posed by malicious MCP servers. 
Mitigating these threats requires concerted efforts from all parties in the MCP ecosystem to strengthen security.
In \autoref{fig:mcp_security_design}, we illustrate the roles, responsibilities, and possible security mechanisms of each stakeholder involved, with further details discussed below.

\textbf{MCP server platforms: pre-release auditing and post-release monitoring.}
MCP server platforms should serve as the first line of defense against malicious MCP servers. 
They are responsible for auditing uploaded servers prior to release and for monitoring published servers. 
For the auditing procedure, first, static and dynamic analyses should be applied to detect malicious logic in server source files and unsafe parameters in server configurations. 
Second, natural-language content such as server metadata, tool descriptions, and resource outputs should be examined for misleading or manipulative language. 
Finally, platforms must verify consistency between a server’s declared behavior and its actual implementation.

\textbf{MCP host: configuration validation, trust management, runtime supervision, and transparency.}
The host application plays a central role in MCP architecture, as it mediates all communications between servers, LLMs, and users. 
Its responsibilities include:

\textit{(1) Safe configuration handling.} Before the host launches or connects to a server, it should inspect the registered configuration block. If over-privileged parameters or unsafe URLs are detected, the host should refuse to launch the server.

\textit{(2) Trust management.} The host could maintain an allowlist of verified servers and a denylist of known malicious ones. 
A tiered trust model may also be adopted. For instance, community servers may be assigned medium trust, while official or signed servers are granted high trust.

\textit{(3) Runtime inspection and filtering.} 
The host should perform inspections when the client receives or sends messages. 
For incoming content, coercive language or malicious binary should be blocked. 
For outgoing content, the host should check whether sensitive data or personal information is being transmitted.

\textit{(4) Safety system prompts.} 
As shown in our experiments, different system prompts can significantly affect the attack outcome. 
Hosts can append safety-oriented system prompts to remind the LLM of MCP-related risks and to reinforce safe interaction principles.

\textit{(5) Tool invocation constraints.} 
Since the LLM invokes tools through the client, the host should enforce control and oversight. 
This includes restricting excessive or context-inappropriate tool calls (\textit{e.g.}, file inspection tasks should not trigger web access tool) and requiring explicit user approval before invoking certain tools.

\textit{(6) Transparency.} 
The host should display the full content of server prompts and resources to users and remind them to verify the content before it is forwarded to the LLM. 
In addition, the host should ensure that all host–server interactions are transparent to the user, including each tool invocation along with its inputs and outputs.

\textbf{LLM: MCP security-oriented training.}
The LLM should undergo targeted training to develop the following capabilities:  
(1) Treat server-provided data with caution rather than blind trust.  
(2) Distinguish between user instructions and server outputs, maintain the primary goal of fulfilling user queries, and resist malicious instructions embedded in server outputs.  
(3) Act as a safety barrier between servers and the user; even if a server returns malicious content, the LLM should not deliver it to the user.  

\textbf{User: awareness and safety practice.}
Users must remain aware of the risks posed by MCP servers.  
When installing third-party servers, they should exercise caution and avoid blindly trusting those that appear attractive.  
For locally installed servers, users should inspect the source code and restrict their privileges, for example by launching them with a user account that has the least privilege.

The safeguard mechanisms we have outlined are not exhaustive.
Yet even these basic security practices remain largely unimplemented in today’s MCP ecosystem. 
More critically, the boundaries of responsibility among stakeholders are still unclear, leaving significant gaps in security governance. 
We therefore call for the establishment of clear security policies and well-defined responsibilities across registries, hosts, LLM providers, and end users. 
Only through collective effort can we move toward a secure and trustworthy MCP-based agent ecosystem.

\section{Conclusion}
\label{sec:conclusion}

In this paper, we conducted a systematic study of the security threats posed by malicious MCP servers.
By analyzing MCP servers from the perspective of their components, we proposed a comprehensive taxonomy that categorizes 12 classes of attacks targeting the local environment, MCP hosts, and users.
Through extensive evaluations across different host–LLM combinations, we demonstrated that these attacks are not only theoretically possible but also highly effective in practice.
Moreover, our investigation revealed that such attacks can be implemented with low effort, while existing detectors remain insufficient against them.
Our findings highlight that malicious MCP servers constitute a primary threat to the rapidly expanding MCP ecosystem.
Addressing this challenge requires more than incremental defenses; it calls for coordinated efforts involving protocol designers, host developers, LLM providers, and the wider community.

\bibliographystyle{ACM-Reference-Format}
\bibliography{main}

\end{document}